\documentclass{elsart}

\usepackage{graphicx}
\usepackage{amssymb}
\usepackage{epsf}

\begin{document}
\begin{frontmatter}

\journal{SCES'2001: Version 1}

\title{\boldmath Slow Spin Dynamics in Non-Fermi-Liquid UCu$_{5-x}$Pd$_x$, $x = 1.0$ and 1.5}

\author[UCR]{D.~E. MacLaughlin\corauthref{1}}
\author[LANL]{R.~H. Heffner}
\author[CSULA]{O.~O. Bernal}
\author[KOL]{G.~J. Nieuwenhuys}
\author[LANL]{J.~E. Sonier\thanksref{SFU}}
\author[UCR]{M.~S. Rose}
\author[UCSD]{R. Chau\thanksref{LLNL}}
\author[UCSD]{M.~B. Maple}
\author[UF]{B. Andraka}

\address[UCR]{University of California, Riverside, California 92521, U.S.A.}
\address[LANL]{Los Alamos National Laboratory, Los Alamos, New Mexico 87545, U.S.A.}
\address[CSULA]{California State University, Los Angeles, California 90032, U.S.A.}
\address[KOL]{Kamerlingh Onnes Laboratory, Leiden University, 2300 RA Leiden, The Netherlands}
\address[UCSD]{University of California, San Diego, La Jolla, California 92093, U.S.A.}
\address[UF]{University of Florida, Gainesville, Florida 32611, U.S.A.}

\thanks[SFU]{Present address: Simon Fraser University, Burnaby, British Columbia, Canada V5A 1S6.}
\thanks[LLNL]{Present address: Lawrence Livermore National Laboratory, Livermore, California 94550, U.S.A.}

\corauth[1]{Corresponding Author: Department of Physics, University of California, Riverside, CA 92521-0413, U.S.A. Phone: (909) 787-5344 Fax: (909) 787-4529 E-mail: douglas.maclaughlin@ucr.edu}

\begin{abstract}

Low-temperature muon spin-lattice relaxation measurements in the non-Fermi-liquid heavy-fermion alloys UCu$_{5-x}$Pd$_x$, $x = 1.0$ and 1.5, indicate inhomogeneously distributed {\em f\/}-electron spin fluctuation rates, and exhibit a time-field scaling of the muon relaxation function indicative of long-lived spin correlations. In UCu$_4$Pd the scaling exponent~$\gamma$ is small and temperature independent. In UCu$_{3.5}$Pd$_{1.5}$ $\gamma$ varies with temperature, increasing with decreasing temperature similar to spin-glass {\em Ag\/}Mn. This suggests that the spin-glass state found for $x \gtrsim 2$ in UCu$_{5-x}$Pd$_x$ modifies the low-frequency spin dynamics in UCu$_{3.5}$Pd$_{1.5}$.

\end{abstract}


\end{frontmatter}

In a number of $f$-electron heavy-fermion compounds and alloys deviations of low-temperature thermal and transport properties from predictions of Landau Fermi-liquid theory signal significant modification of the low-lying excitations of these systems~\cite{ITP96}. There is strong evidence, primarily from nuclear magnetic resonance~\cite{MacL00} and muon spin rotation~\cite{MHSN00} experiments, that disorder can play an important role in this non-Fermi liquid (NFL) behavior. 

This paper reports muon spin-lattice relaxation measurements in the NFL alloys UCu$_{5-x}$Pd$_x$, $x = 1.0$ and 1.5 (see also Ref.~\cite{MBHN01}). A characteristic {\em time-field scaling\/} of the muon asymmetry relaxation function~$G(t,H) = G(t/H^\gamma)$ is observed, which shows that these systems exhibit long-lived spin correlations reminiscent of quantum critical dynamics~\cite{AORL95b} or of disordered materials such as spin glasses~\cite{KMCL96,PSAA84}.

In a system with a spatially inhomogeneous distribution of local muon relaxation rates~$W({\bf r},H)$, the sample-average muon asymmetry relaxation function~$\overline{G}(t,H)$ is given by
\begin{equation}
\overline{G}(t,H) \propto \int \d{\bf r}\, \exp[-W({\bf r},H)t] \,,
\end{equation}
which yields a nonexponential time dependence. In turn, $W({\bf r},H)$ is related to the imaginary component~$\chi''({\bf r},\omega_\mu)$ of the local dynamic spin susceptibility:
\begin{equation}
W({\bf r}, H) \propto \chi''({\bf r},\omega_\mu)/\omega_\mu \,,
\end{equation} 
$\omega_\mu = \gamma_\mu H$. The noise power of the thermal spin fluctuations at the muon Zeeman frequency~$\omega_\mu$ gives rise to the magnetic field dependence of $\overline{G}(t,H)$, and a strong field dependence indicates slow spin fluctuations with a correlation function characterized by a long-time ``tail''~\cite{MBHN01,KMCL96}.
 
\begin{figure}[t]
\begin{center} \leavevmode
\epsfxsize=12cm
\epsfbox{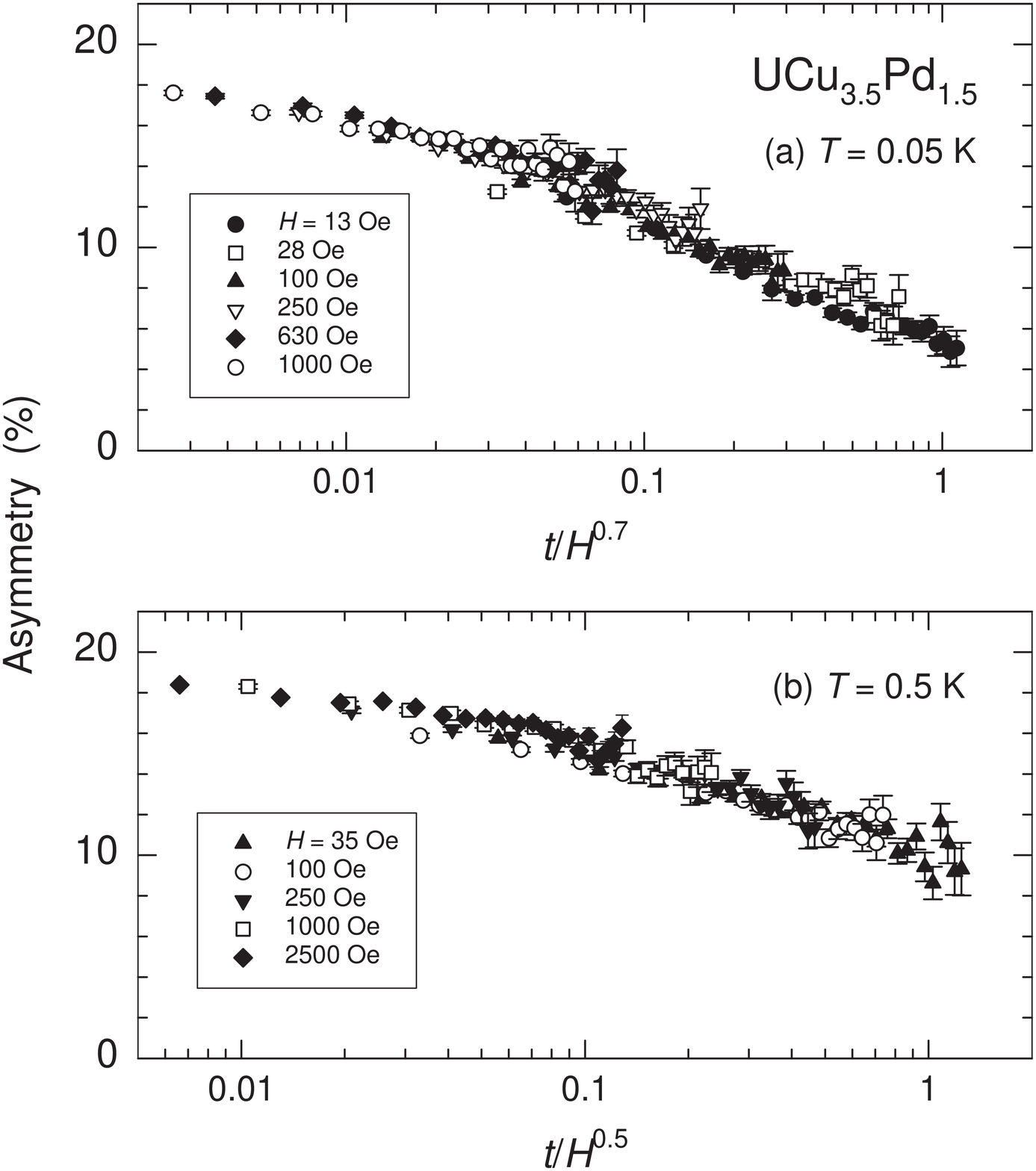}
\caption{Dependence of sample-average muon asymmetry relaxation 
function~$\overline{G}(t)$ on scaling variable~$t/H^\gamma$ in UCu$_{3.5}$Pd$_{1.5}$. (a)~$T = 0.05$~K, $\gamma = 0.7\pm0.1$. (b)~$T = 0.5$~K, $\gamma = 0.5\pm0.1$.}
\label{fig:UC35Pscaling}
\end{center}
\end{figure}

\begin{figure}[t]
\begin{center} \leavevmode
\epsfxsize=12cm
\epsfbox{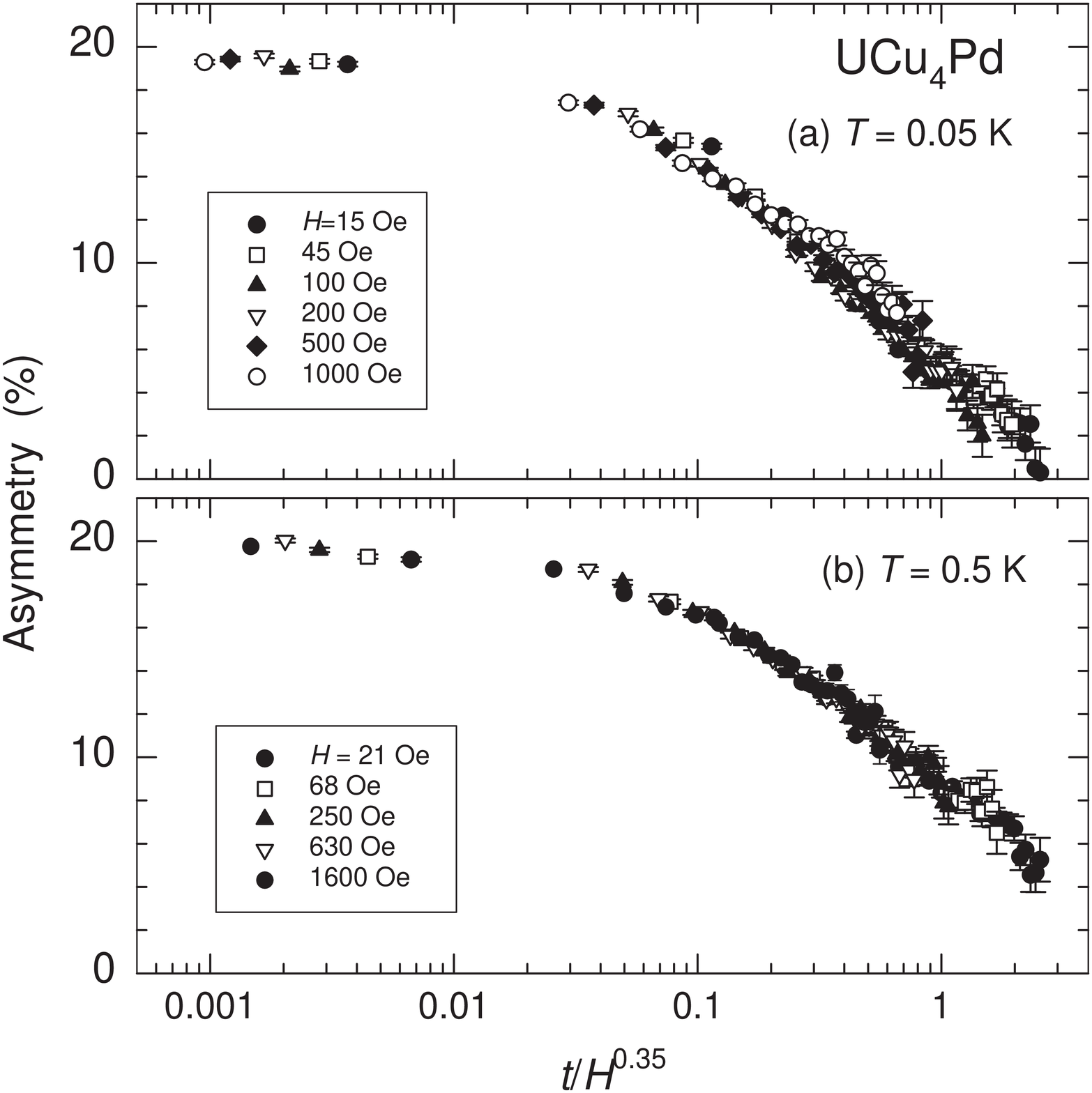}
\caption{Dependence of $\overline{G}(t)$ on $t/H^\gamma$ in UCu$_4$Pd. (a)~$T = 0.05$~K\@. (b)~$T = 0.5$~K\@. At both temperatures $\gamma = 0.35 \pm 0.10$.}
\label{fig:UC4Pscaling}
\end{center}
\end{figure}

$\mu$SR asymmetry data for UCu$_{3.5}$Pd$_{1.5}$ and UCu$_4$Pd are given in Figs.~\ref{fig:UC35Pscaling} and \ref{fig:UC4Pscaling}, respectively, for temperatures of 0.05~K and 0.5~K\@. The data are plotted as functions of the scaling variable~$t/H^\gamma$, where $\gamma$ was varied to obtain the best scaling, i.e., the most nearly universal behavior. The observation of time-field scaling indicates that the local spin dynamics are characterized by a dynamic susceptibility which obeys a power law
\begin{equation}
\chi''({\bf r},\omega)/\omega \propto \omega^{-\gamma}
\end{equation}
at low frequencies. It is important to note that no particular form for $\overline{G}(t)$ is assumed in this analysis. There is some indication that for $T = 0.05$~K scaling begins to break down at $\mbox{fields} \gtrsim 1000$~Oe ($\mu_BH \gtrsim k_BT$); it is not surprising that fields of this magnitude would begin to affect spin dynamics. No loss of scaling is observed at 0.5~K for fields as large as 2500~Oe.

In UCu$_{3.5}$Pd$_{1.5}$ $\gamma$ from muon relaxation exhibits a temperature dependence, becoming larger at low temperatures: $\gamma(0.5~{\rm K}) = 0.5\pm0.1;\ \gamma(0.05~{\rm K}) = 0.7\pm0.1$. This is reminiscent of the behavior of spin-glass {\em Ag\/}Mn above the spin-freezing ``glass'' temperature~$T_g$, where $\gamma$ increases as $T \rightarrow T_g$~\cite{KMCL96}. It suggests a continuous temperature dependence of $\gamma$ in UCu$_{3.5}$Pd$_{1.5}$ between the low-temperature muon relaxation results and the higher-temperature ($> 10$~K) inelastic neutron scattering (INS) value~$\gamma = 0.33$~\cite{AORL95b}. 

In UCu$_4$Pd, on the other hand, $\gamma = 0.35\pm0.10$ from muon relaxation at both 0.05~K and 0.5~K (Fig.~\ref{fig:UC4Pscaling}). This value is smaller than in UCu$_{3.5}$Pd$_{1.5}$, and in agreement with INS experiments above 10~K~\cite{AORL95b} which give $\gamma = 0.33$ as in UCu$_{3.5}$Pd$_{1.5}$. This temperature independence suggests that in UCu$_4$Pd the slow fluctuations are quantum rather than thermal in origin. In UCu$_{3.5}$Pd$_{1.5}$ the Pd concentration~$x$ is closer to the value $x \approx 2$ for which a spin-glass phase occurs~\cite{AnSt93}, and the proximity of this phase may be reflected in the low-temperature increase of $\gamma$.

Time-field scaling and other aspects of these experiments are discussed in detail elsewhere~\cite{MBHN01}.

We are grateful to W.~P. Beyermann, A.~H. Castro Neto, and L.~P. Pryadko for discussions of these issues, and to A. Amato, C. Baines, and D. Herlach for help with the experiments. This research was supported in part by the U.S. NSF, Grants DMR-9731361 (U.C. Riverside), DMR-9820631 (CSU Los Angeles), and DMR-9705454 (U.C. San Diego), and by the Netherlands NWO and FOM, and was performed in part under the auspices of the U.S. DOE.

\end{document}